      [1999/12/01 v1.4c Il Nuovo Cimento]
\renewcommand{\thefootnote}{\fnsymbol{footnote}}
\documentclass{cimento}

\title{GRBs Optical follow-up observation at Lulin observatory, Taiwan}
\author{K.Y. Huang\from{1}\ETC,
Y. Urata\from{2} \from{3},
W.H. Ip\from{1},
T. Tamagawa\from{2},
K. Onda\from{2}
        \atque
K. Makishima\from{2}\from{4} 
}
\instlist{
  \inst{1}Institute of Astronomy, National Central University,
  Chung-Li 32054, Taiwan, Republic of China
  \inst{2}RIKEN (Institute
  of Physical and Chemical Research), 2-1 Hirosawa, Wako, Saitama
  351-0198, Japan 
  \inst{3}Department of Physics, Tokyo Institute of
  Technology, 2-12-1 Oookayama, Meguro-ku, Tokyo 152-8551, Japan
  \inst{4}Department of Physics, University of Tokyo, 7-3-1 Hongo, Bunkyo-ku, Tokyo 113-0033, Japan

}

\PACSes{\PACSit{95.55.Cs}{Ground-based ultraviolet, optical and infrared telescopes} 
\PACSit{98.70.Rz}{ gamma-ray source; gamma-ray bursts}}
\begin{document}

\maketitle

\begin{abstract}
   The Lulin GRB program, using the Lulin One-meter Telescope (LOT) in
 Taiwan started in July 2003. Its scientific aims are to discover
 optical counterparts of XRFs and short and long GRBs, then to quickly
 observe them in multiple bands. Thirteen follow-up observations were
 provided by LOT between July 2003 and Feb. 2005. One host galaxy was
 found at GRB 031203. Two optical afterglows were detected for GRB
 040924 and GRB 041006. In addition, the optical observations of GRB
 031203 and a discussion of the non-detection of the optical afterglow
 of GRB 031203 are also reported in this article.
   
\end{abstract}

\section{Introduction}
 
  A parallel effort based on the Kiso GRB optical observation system
\cite{14,15}, was started in July 2003 using the Lulin One-meter
Telescope (LOT) at Taiwan.  The scientific aims of our Lulin GRB
program are: (1) to discover optical counterparts of XRFs, and short-
and long-duration GRBs; (2) to perform multi-band observations that
will unlock the temporal and spectral evolution of the corresponding
optical afterglows.  The Lulin observatory is located in Nantou
County, Taiwan, at $120^\circ52'25''\it{E}$, $23^\circ28' 07''\it{N}$,
on a 2862$-$m high peak. The sky background levels in the $UBVRI$
bands are : $U$ = 21.8; $B$ = 22.0; $V$ = 21.3; $R$ = 20.9 ; $I$ =
19.5 $\it{mag}$ $\it{arcsec^{-2}}$, respectively.  From statistical
measurements at Lulin site, it was found that the typical accumulative
observation time is 1700 hours per year and the typical seeing is 1.5
$\it{arcsec}$ .  The LOT is a Cassegrain telescope (f/8) with an ACE
filter system and a main CCD camera -- PI1300B (1340 $\times$ 1300,
F.O.V.$\sim$ 11.5$^{\prime}$ $\times$ 11.2$^{\prime}$) as well as a
spare CCD camera -- Ap8 (1k $\times$ 1k, F.O.V. $\sim$
10.6$^{\prime}$$\times$ 10.6$^{\prime}$). Table I shown the
characteristics of LOT with AP8 and PI1300B CCD \cite{1}.

\begin{table}
  \caption{Basic characteristics of LOT with PI1300B and AP8 CCD}
\begin{center}
\vspace{-0.4cm}
\begin{tabular}{ccccccc} \hline
       &                & PI1300B CCD      &                 &                &    AP8 CCD       &            \\ \hline
Filter & Zero point \small{*}&   Color term & Extinction      & Zero point    &   Color term     & Extinction \\ \hline
 B     & 22.34$\pm$0.02 & 0.199$\pm$0.018  &  0.19$\pm$0.02  & 21.99$\pm$0.03 & 0.035$\pm$0.005  &  0.28$\pm$0.02   \\  
 V     & 22.68$\pm$0.02 & $-$0.058$\pm$0.017  &  0.11$\pm$0.01  & 21.97$\pm$0.02 & 0.069$\pm$0.004  &  0.20$\pm$0.02  \\     
 R     & 22.66$\pm$0.01 & $-$0.049$\pm$0.021  &  0.09$\pm$0.01   & 21.87$\pm$0.02 & 0.113$\pm$0.007  &  0.15$\pm$0.02   \\ 
 I     & 21.99$\pm$0.04 & 0.040$\pm$0.029  &  0.06$\pm$0.01   & 21.27$\pm$0.03 & 0.043$\pm$0.005  &  0.12$\pm$0.02\\ \hline

\end{tabular}
\small{*}{Units : Zero point (mag), Color term (mag), Extinction (mag/airmass)}

\end{center}
\end{table}
\vspace{0.1cm}
  Due to the fact that there are only a very few telescopes for GRB
follow-up observations in East Asia, and that the observational range
can reach up to Dec. $-$40 degree, Lulin enjoys a unique position for
GRB study. The fast PI1300B read out time that allows small cadence
multi-band time series photometry is another distinct advantage. Due
to the nature of GRBs detection, the Lulin GRB project is included as
a part of the Target of Opportunity (TOO) program. We have developed
two approaches to search for GRB optical afterglows, according to the
dimensions of the burst error box provided by such satellites as
$\it{HETE-2}$, $\it{INTEGRAL}$, and $\it{SWIFT}$. First, if the error
range is larger than the FOV of the LOT, we can dither the field and
search for the optical counterpart in the $B$ or $R$ band. Second, if
the error range is smaller than the FOV, we can quickly locate the
counterpart and monitor the temporal variation of the brightness of
the optical afterglow in several wavelengths.

  During the analysis procedure, the positions and coordinates of the
detected objects in our images are compared with the USNO stellar
coordinates. These physical coordinates are then transformed to
equatorial coordinates using the WCS (World Coordinate System). After
this, our images compared with DSS2 images, to look for likely
candidates.

\section{Summary of observed events}

 Between July 2003 and Feb. 2005 (Table II), thirteen follow-up
observations were provided by LOT. We could provide upper limits to
the magnitude for ten events. Two optical afterglows were detected for
GRB 040924 and GRB 041006 and the host galaxy of GRB 031203 was
identified. We will focus on the GRB 031203 case in the following.

\begin{table}
  \caption{Log of Lulin GRB follow-up observations.}
\begin{center}
\vspace{-0.45cm}
\begin{tabular}{ccccccc} \hline
   GRB     & Delay time & Triggered & Limit mag.       & Results     & Publications \\
           &     (hr)   & Spacecraft&  (3$\sigma$)     &             &               \\ \hline  
GRB 030823 &   3.3      &  $\it{HETE-2}$   &  R $\sim$ 19     & Upper limit &  GCN 2360      \\
GRB 031026 &   6.0      &  $\it{HETE-2}$   &  R $\sim$ 21     & Upper limit &  GCN 2436      \\
GRB 031203 &  19.6      & $\it{INTEGRAL}$  &  I $\sim$ 20     & Host Galaxy &  GCN 2470      \\
GRB 031220 &   8.0      & $\it{HETE-2}$   &  R $\sim$ 20     & Upper limit &  GCN 2494      \\
GRB 040422 &   9.6      & $\it{INTEGRAL}$ &  R $\sim$ 20     & Upper limit &  GCN 2577      \\
GRB 040916 &  16.3      &  $\it{HETE-2}$   &  I $\sim$ 20     & Upper limit &  GCN 2721      \\
GRB 040924 &   2.3      &  $\it{HETE-2}$   &      $-$         &    OT       &  GCN 2744      \\
GRB 041006 &   9.0      &  $\it{HETE-2}$   &      $-$         &    OT       &  GCN 2785      \\
GRB 041211 &   3.4      &  $\it{HETE-2}$   &  R $\sim$ 20     & Upper limit &  GCN 2840      \\
GRB 041219 &   9.6      &  $\it{SWIFT}$    &  R $\sim$ 21     & Upper limit &  GCN 2891      \\
GRB 050123 &   5.5      &  $\it{HETE-2}$  &  I $\sim$ 19     & Upper limit &  GCN 2971      \\
GRB 050124 &   4.4      &  $\it{SWIFT}$    &  R $\sim$ 20     & Upper limit &  GCN 2976      \\

GRB 050215B &   13.1      & $\it{SWIFT}$      &  V $\sim$ 19     & Upper limit &  GCN 3030      \\
            &             &             &  I $\sim$ 20    & Upper limit &  GCN 3030 \\
\hline 
\end{tabular}
\end{center}
\end{table}

  GRB 031203 was detected by the IBIS instrument on $\it{INTEGRAL}$ on
December 3, 2003 at 22:01:28 UT as a single peaked burst with a
duration of 30s \cite{2}. $\it{XMM-Newton}$ detected two sources named S1
and S2 within the INTEGRAL error circle. The brightest source S1, which
faded through out the observation period, was interpreted to be the
afterglow \cite{3}. Its position coincided with that of a fading radio
source \cite{4}.


\begin{figure}[tbhp]
\begin{center}
\psfig{figure=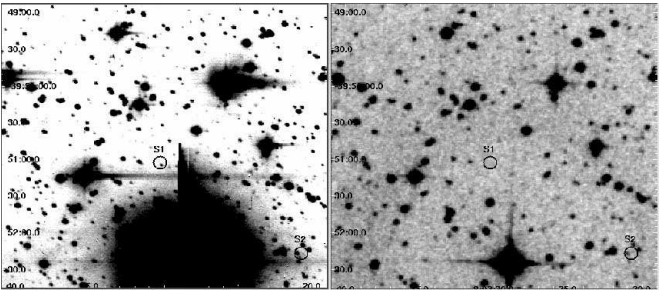,height= 1.5in,angle=270}
\vspace{-0.2cm}
\caption{LOT $I-$band image of $\it{XMM}$-Source : S1 and S2. By comparing this with the DSS $I-$band image, we found a source near S1.}
\end{center}
\end{figure}


 A new source was detected by Lulin $I-$band observations \cite{5}; see
Figure 1. Bloom et al. \cite{6} have pointed out that this new source could
either be a galaxy in the foreground or the host of GRB 031203. A
radio afterglow was detected at the same position. From spectroscopic
observations, Prochaska et al. \cite{7} found this to be an active
star-forming galaxy, with $z = 0.105$ and that should be the host
galaxy of GRB 031203, designated HG 031203.

   The LOT follow-up observations in combination with other I-band
 data \cite{8,9,10}, indicated that the source did not show variability in
 terms of its brightness (Figure 2). Thus, the optical source detected
 was the host galaxy, HG 031203, no variable source brighter than $I$
 = 20.0 was found during our observations. The astrometry of the GRB
 031203 field was corrected by matching with stars in the USNO A2
 catalog. This resulted in the following astrometric position of HG
 031203: R.A. = $08^{\rm h} 02^{\rm m} 30^{\rm s}.177$ $\pm$ 0.175
 $\it{arcsec}$, Dec. = $-39^{\circ} 51' 03''.960$ $\pm$ 0.164
 $\it{arcsec}$ (J2000.0). It is important to note that optical and
 spectroscopic observations made by several groups have revealed that
 SN 2003lw was associated with GRB 031203. It had a brightness peak
 between 26 and 34 days after the gamma-ray burst \cite{8,9,10,11}.

\begin{figure}[htb]
\vspace{-0.6cm}
\psfig{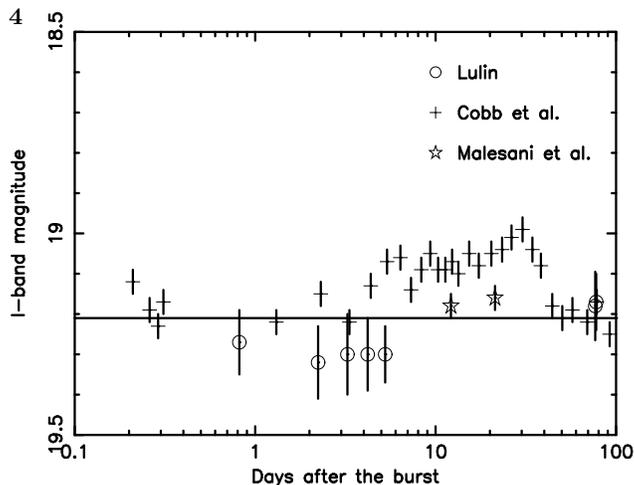}
\vspace{-0.3cm}
\caption{Lightcurve of HG 031203 at the $I-$band. The solid line shows the magnitude of the host galaxy ($I$ = 19.21).}
\end{figure} 

  In order to explore the reason behind the absence of optical
afterglow, we have analyzed the X-ray $\it{XMM-Newton}$ afterglow
data. Using the power law plus absorption model, we found that the
best fit of the photon index is 1.8$\pm$0.1 ($\chi^2$=14.07 with 22
dof) with $N_H$ = 7.73${\pm}$0.04$\times 10^{21} cm^{-2}$ for the
energy between 0.6 keV to 10 keV. Our results are consistent with
those found by Watson et al. \cite{12}. Assuming no break in the
lightcurve, the spectral flux distribution follows ${F_{\nu}}$
$\propto$ ${\nu^{-\beta}}$ ($\beta = 0.8\pm0.1$) and the extrapolation
of the X-ray flux to the optical range yields $I$ = 20.65 and $R$ =
21.13, at 0.26 days after the burst.

  In comparison with the value of $N_H$ = 7.73${\pm}$0.04$\times
10^{21} \it{cm^{-2}}$, the neutral hydrogen column density along the
line-of-sight to GRB 031203, is on the order of $N_H$ = 6.21 $\times
10^{21} \it{cm^{-2}}$ within the Galaxy \cite{13}. This means that most of
the extinction of the optical emission of this gamma-ray burst should
have come from the Galactic interstellar medium. This effect alone is
enough to reduce the optical brightness by 2 magnitudes from 20.65 to
about 22.65 at the $I-$band \cite{13}. This might explain why GRB 031203
shows no optical afterglow.

\vspace{-0.35cm} 
\acknowledgments
We thank all those of  Lulin users for helping with the GRB follow-up
observations.  This work is supported by NSC 93-2752-M-008-001-PAE and
NSC 93-2112-M-008-006. K.Y. Huang acknowledges support from from
Foundation For the Advancement of Outstanding Scholarship.

\end{document}